\documentstyle[12pt]{article}
\begin{document}
\def\be{\begin {equation}}
\def\ee{\end {equation}}
\def\ba{\begin {equnarray}}
\def\ea{\end {equanarray}}
\def\m{\mathrm}

\begin{center}
{\Large \bf On the Effective Electron Mass in Magnetar-like Fields} \\
\vspace {1.0 cm}
M. Conte$^1$, M. Palazzi$^1$, M. Pusterla$^2$ and R. Turolla$^2$. \\
\vspace {0.5 cm}
{\small \it
(1) Dipartimento di Fisica dell'Universit\`{a} di Genova, INFN Sezione di
Genova, \\
Via Dodecaneso 33, 16146 Genova, Italy. \\
(2) Dipartimento di Fisica dell'Universit\`{a} di Padova, INFN Sezione di
Padova, \\
Via Marzolo 8, 35131 Padova, Italy.}
\end {center}

\vspace {1.0 cm}

\begin{abstract}
We show, either quantum mechanically or classically, that the variation of the
effective mass induced in a charged particle by the presence of an ultra-strong
electromagnetic field may lead to observable consequences. In particular, we discuss
how this effect may operate on electrons close to the surface of a magnetar, a
neutron star with a super-strong magnetic field of the order of $10^{11}$ Tesla.
\end{abstract}

\section {Quantum Mechanical Approach}

$~~~~$ The relativistic mass of a point particle is modified  by additional terms
which arise from its interaction with a given external scalar, vector or tensor
field. However, the mass difference due to this renormalization is irrelevant for
present man-made magnetic fields, of the order of few tens of Tesla, at most.
Nevertheless it may become relevant in astrophysical objects in peculiar situation
for electrons, namely when these particles move within extremely high magnetic
fields that seem to be present in special neutron stars, called {\bf magnetars}
\cite{NS1},\cite{NS2}, which are characterized by magnetic fields of the order of
$10^{11}$ Tesla. To be more explicit, the effective mass that the electron acquires
in such ultra-high fields, is expected to generate observable effects. In order to
evaluate the amount of these effects, we shall make use \cite{KU} of the Green
function $G(x,y)$ for a Dirac particle in a homogeneous magnetic field,
characterized by a potential $A_\nu$; the related equation is

\be \left(i\;\frac {\partial}{\partial x^\nu}\;\gamma^\nu - \frac {e}{\hbar}\;
    A_\nu\gamma^\nu - \kappa\right)S^e(x,y) = \delta^4(x-y)
       ~~~ {\m
       with}  ~~~ \kappa = \frac {mc}{\hbar}  \label {eq1} \ee
Setting

\be S^e(x,y) = \left(i\;\frac {\partial}{\partial x^\mu}\;\gamma^\mu -
          \frac{e}{\hbar}\;A_\mu\gamma^\mu + \kappa\right)G(x,y)    \label {set} \ee
eq. (\ref{eq1}) transforms into

\be \left(-\frac{\partial}{\partial x^\nu}\frac{\partial}{\partial x_\nu} +
    \frac {e^2}{\hbar^2}\;A_\nu A^\nu -
    \frac {ie}{\hbar}\;\frac{\partial A_\mu}{\partial x^\nu}\;\gamma^\nu\gamma^\mu -
    \frac {2ie}{\hbar}\;A^\nu\;\frac{\partial}{\partial x^\nu}\right)G(x,y) =
                                               \delta^{(4)}(x-y)    \label {eq2} \ee
Choosing the reference frame with the $z$-axis parallel to the uniform magnetic
field $\vec B$, we have

\be \cases {A_1 = A_x = -\frac{1}{2}Bx_2 = -\frac{1}{2}By \cr
            A_2 = A_y = -\frac{1}{2}Bx_1 = -\frac{1}{2}Bx \cr
            A_3 = A_z = 0 \cr
            A_0 = 0}                                        \label {pot} \ee
In the impulses space, the Green function can be written as

\be G(r,s) = \frac {1}{(2\pi)^2}\int e^{irx} e^{-isy} G(x,y) d^4x d^4y
                                                            \label {G1} \ee
where $rx$ and $sy$ are the internal products between the wave tetra-vectors

\be r^\mu = \frac {p^\mu}{\hbar} ~~~ {\m and} ~~~ s^\mu = \frac {q^\mu}{\hbar}
                                                             \label {4wv} \ee
with $p^\mu$ and $q^\mu$ initial and final tetra-momentum respectively, and the
tetra-vectors $x$,$y$, while $d^4x$ and $d^4y$ govern the integration in the four
dimensions space. Moreover, by making use of the Whittaker functions W, eq.
(\ref{G1}) transforms into

\be G(r,s) = f(r,s)\left(
\begin{array}{cc}
\Gamma (1+a/b)W_{-1/2-a/b,0}(\zeta) & 0 \\
0 & \Gamma (a,b)W_{1/2-a/b,0}(\zeta)
\end{array}\right)
\label {G3} \ee
with

\be f(r,s) = \frac {1}{32w^2\pi^4}\delta(r_0-s_0)e^{\{(i/w)(r_1s_2-s_1r_2)\}}
             \zeta^{-1/2-a/2b}                                  \label {frs} \ee
and

\be \zeta = \frac {1}{w}[(r_1-s_1)^2 + (r_2-s_2)^2], ~~
        w = \frac {eB}{2\hbar}, ~~
        a = r_0^2 - r_3^2 - \kappa^2, ~~
        b = -4w                                            \label {param} \ee
The functions $\Gamma(a/b)$ and $\Gamma(1+a/b)$ have zeroes for

\be  a/b = -n ~~~~~~~~~~~ (n = {\m integer} >0)                \label{p1} \ee
then, choosing $r_3=0$, we have

\be \frac {r_0^2-\kappa^2}{-\frac {4eB}{2\hbar}} = -n ~~ \Longrightarrow ~~
    r_0^2 - \frac {m^2c^2}{\hbar^2} = n \frac {2eB}{\hbar} ~~ \Longrightarrow ~~
     \frac {E^2}{c^2\hbar^2} = \frac {m^2c^2}{\hbar^2} + n \frac {2eB}{\hbar}
                                                            \label {p2} \ee
or

\be E = m^*c^2 = mc^2\sqrt{1 + \frac {2neB\hbar}{m^2c^2}}       \label {meff1} \ee

 Take note that the most important poles are $n=1$ and $n=2$ since higher values are
practically cancelled by the factorials. As it will be justified in the next
Section, the second term inside the square root is modified as follows:

\be \frac {2neB\hbar}{m^2c^2} = 2nB \frac {2}{mc^2} \frac {e\hbar}{2m} \simeq
    2nB \frac {2\mu}{mc^2} = 2n\;\frac {B}{B_{\m Q}}            \label {mod1} \ee
since in the $e^\pm$ example, the magnetic moment is

\be \mu = g\frac {e\hbar}{4m} \simeq \frac {e\hbar}{2m}      \label {mue} \ee
Moreover, the quantity

\be B_{\m Q} = \frac {mc^2}{2\mu} = 4.41 \times 10^9 ~ T     \label {BQ} \ee
can be considered as that magnetic field which provokes a spin-flip characterized by
an energy equal to the rest-mass energy of the involved particle, i.e. $2B\mu=mc^2$.
Hence, combining eqs. (\ref{meff1}) and (\ref{mod1}), we obtain

\be m^* = m\sqrt{1+2n\frac{B}{B_{\m Q}}} = m\sqrt{46.35} = 6.81\;m
                                                             \label {meff2} \ee
for $n=1$ and $B=10^{11}$ Tesla as typical for magnetars. In {\bf Table I} we quote
a few data of interest regarding both electrons an protons, the latter mentioned for
demonstrating how smaller is their influence on the effective mass.

\begin {center}
{\bf Table I}
\par\vspace {0.25 cm}
\begin {tabular} {|c|c|c|}
\hline
       Data      &    $e^\pm$               &    $p(\bar p)$       \\
\hline \hline
 $mc^2~(J)$      & $8.19 \times 10^{-14}$  & $1.50 \times 10^{-10}$ \\
\hline
 $\mu~(JT^{-1})$ & $9.28 \times 10^{-24}$  & $1.41 \times 10^{-26}$ \\
\hline
 $B_{\m Q}~(T)$  & $4.41 \times 10^{9}$    & $5.33 \times 10^{15}$ \\
\hline
\end {tabular}
\end {center}
\vspace {0.25 cm}
\par\noindent

\section {Classical Approach}

$~~~~$ In this section we consider \cite{Barut} a mass point charged particle that
has an intrinsic magnetic moment $\vec \mu$ and intrinsic electric dipole $\vec d$
in presence of an electromagnetic external field $\vec B$, $\vec E$. As usual, the
interaction with $\vec \mu$ and $\vec d$ is

\be
 \frac{1}{2}\sigma^{\mu \nu} F_{\mu \nu} =
 (\vec \mu \cdot \vec B + \vec d \cdot \vec E)
\label{sigf} \ee
where $\sigma^{\mu \nu}$ is an antisymmetric tensor built with the components of
$\vec \mu$ and $\vec d$ in the particle rest frame.

\be \sigma^{\mu \nu}= \left(
\begin {array}{cccc}
0 & -cd_1 & -cd_2 & -cd_3 \\
cd_1 & 0 & \mu_3 & -\mu_2 \\
cd_2 & -\mu_3 & 0 & \mu_1 \\
cd_3 & \mu_2 & -\mu_1 & 0 \\
\end {array}
\right) \label{sigma} \ee
and $F^{\mu \nu}$ is the electromagnetic tensor. Then in the laboratory frame we
have:

\be \cases { \vec \mu_{lab}=\gamma \left[\vec \mu - \frac{\gamma}{\gamma+1} ( \vec
\mu \cdot \vec \beta ) \vec \beta \right]-\gamma c  \left( \vec \beta \times \vec d
\right) \cr \vec d_{lab}=\gamma \left[ \vec d -\frac{\gamma}{\gamma+1} \left(  \vec
d \cdot \vec \beta \right) \vec \beta \right] + \gamma \left( \frac{\vec \beta} {c}
\times\vec \mu \right)}      \label{diplab} \ee
Let us consider the corresponding action in the covariant \cite{Feynman} form

\be S=\int L'dt=\int L \left( x^\mu,~\frac{d x^\mu}{d \alpha};~\alpha \right)
d\alpha          \label{action} \ee
where $\alpha$ is a parameter. Assuming the intrinsic moments as constant, the
Lagrange function becomes:

\be L=L_{em}+L_{int}=\left[ m c \left( \frac{d x^\lambda}{d \alpha} \cdot \frac{d
x_\lambda} {d \alpha} \right) ^{\frac{1}{2}}+ e A_\mu \frac{d x^\mu} {d \alpha}
\right]+ \left[\frac{1}{2c}\sigma^{\mu \nu} F_{\mu \nu}\left( \frac{d x^\lambda}
{d\alpha} \cdot \frac{d x_\lambda}{d \alpha} \right)^{\frac{1}{2}} \right]
\label{lagr} \ee
By considering $\alpha$ coincident with the proper time $\tau$, we can easily derive
from eqs. (\ref{action}) and (\ref{lagr}) the equation \cite{Barut} of motion

\be \frac{d}{d \tau}p_\mu = \frac{d}{d\tau} (m u_\mu) = \frac {1}{1+\Delta} \left
(eF_{\mu \nu} u^{\nu} + \frac{1}{2} \sigma^{\alpha \beta} F_{\alpha\beta,\mu}-
\frac{1}{2} \sigma^{\alpha\beta} {F_{\alpha\beta,\nu}} u^\nu u_\mu \right)
\label{Euler} \ee
under the constraints:

\be \frac {dx^\lambda}{d\tau} \cdot \frac {dx_\lambda}{d\tau} = c^2
                  ~~~~~~~~~ {\m and} ~~~~~~~~~
    \frac {dx^\lambda}{d\tau} \cdot \frac {d^2 x_\lambda}{d\tau^2} = 0  \label {rul}
\ee
Therefore, it is trivial to deduce that the electron rest mass $m$ is modified in
the following way

\be m^* = (1+\Delta)m       \label {meff3} \ee

  The symbol $\Delta$ in eq. (\ref{Euler}) represents a finite term of the mass
renormalization, generated by the coupling between the electromagnetic field and the
intrinsic moments, which in our case becomes

\be \Delta = \frac{1}{2} \frac{\sigma^{\alpha \beta} F_{\alpha \beta}}{m_0 c^2} =
\frac{\vec \mu \cdot \vec B}{m_0 c^2} = \pm \frac {B}{2B_{\m Q}}
                                                  \label {deltam} \ee
where $\pm$ refers to the two possible alignments between a particle of spin 1/2 and
the field lines.

 In fact, for
such fields and bearing in mind the definition (\ref{BQ}), the correcting term
(\ref{deltam}) and the effective mass (\ref{meff3}) become respectively

\be \Delta = 11.34  ~~~~~~~~~~~~~~ {\m and} ~~~~~~~~~~~~~~ m^* = 12.34 \; m
                                                           \label {meff4} \ee
not so different than the value (\ref{meff2}) found via quantum mechanical
considerations.

\section {Astrophysical Implications}

$~~~~$ The electrons move freely in the neutron star atmosphere along the magnetic
field lines, but their transverse motion is quantized and the corresponding energy
levels, dubbed Landau levels, follow the harmonic oscillator formula

\be E_n = \left(n+\frac{1}{2}\right) \hbar\omega_c,  ~~~~ {\m with} ~~~~
                \omega_c = \frac {eB}{m}            \label {landau} \ee

   Transitions between two nearby levels determine the emission, or absorption, of
photons of energy $E_c=\hbar\omega_c$; the detection of these lines represents a
powerful diagnostics which provides an accurate measure of $B$.

   If the real rest mass $m$ of the particles, electrons in our case, is replaced
by its effective mass $m^*$, as given by eq. (\ref{meff2}), one obtains a change of
the cyclotron energy levels accordingly:

\be E_c = \frac {\hbar e B}{m^*} =
 E_c^{(0)}\left[1+2n\frac {B}{B_{\m Q}}\right]^{-\frac{1}{2}},
 ~~ {\m with} ~~ E_c^{(0)} = \frac {\hbar e B}{m} \simeq 10~MeV
 ~~ {\m for} ~~ B \simeq 10^{11}~T
 \label {qjmp2} \ee

   The values of $E_c \simeq 10~MeV$ appear to high for the spectroscopical probes
of the existing space missions, but in magnetic fields of the order of
$10^9\div10^{11}$ Tesla $E_c$ goes down to $\approx 1.5 \ MeV$ i.e. enters into an
energy interval which allows its detection. At this point, we like to recall that
magnetars are considered responsible for the X/$\gamma$-ray emission of
soft-$\gamma$-repeaters (SGRs), and possibly of the anomalous X-ray pulsars (AXPs);
nowadays these sources are very few: about ten. We also remind that, despite of the
fact that about one out of ten supernova events seems to give birth to a magnetar,
the magnetar stage is quite short because of the field decays very rapidly (see e.g.
\cite{Tho} for a review).

   Another important aspect, related with magnetars, appears from the detection \cite{Ibr}
of a proton cyclotron line in the X-ray spectrum of SGR 1806-20 in outburst: it gave
a direct measure of the magnetic field strength for the first time, namely a field
of the order of $10^{11}$ Tesla, thus confirming that SGRs  are magnetars.

   From all the features analyzed above, and bearing in mind that these ultra-high
magnetic fields cab be even slightly bigger than $10^{11}$ Tesla, we may conclude
that, although the mentioned effective mass is negligible for protons, even in the
magnetar magnetic field, the simultaneous detections of a proton line in the $keV$
range, and of an electron line in the 0.5-1 $MeV$ band, are capable of providing a
direct confirmation of the effect described in this note, based on classical
electrodynamics.

   Even if not yet feasible at present, accurate measurements of this sort are going
to become possible in the nearest future by using next generation space
instrumentation.

\end {document}